\documentclass[11pt,twoside]{article}

\usepackage{asp2006}
\usepackage{epsf}

\begin{document}

\title{The Highest Redshift Relativistic Jets}

\author{C.~C. Cheung\altaffilmark{1}, \L. Stawarz}
\affil{Kavli Institute for Particle Astrophysics and Cosmology, \\Stanford 
University, Stanford, CA 94305}    

\altaffiltext{1}{Jansky Postdoctoral Fellow of the National Radio
Astronomy Observatory}

\author{A. Siemiginowska, D.~E Harris, D.~A. Schwartz}
\affil{Harvard-Smithsonian Center for Astrophysics, \\
60 Garden St., Cambridge, MA 02138}

\author{J.~F.~C. Wardle, D. Gobeille}
\affil{Physics Department, Brandeis University, Waltham, MA 02454}

\author{N.~P. Lee}
\affil{Institute for Astrophysical Research, Boston University, \\
725 Commonwealth Ave., Boston, MA 02215}

\begin{abstract} We describe our efforts to understand large-scale (10's--100's kpc)
relativistic jet systems through observations of the highest-redshift quasars.  Results
from a VLA survey search for radio jets in $\sim$30 z$>$3.4 quasars are described along
with new $Chandra$ observations of 4 selected targets. \end{abstract}

\section{Why High-redshift Jets?} 

It is now well established that X-ray emission is a common feature of
kiloparsec-scale radio jets \citep[see][for a recent review and the associated
website, http://hea-www.harvard.edu/XJET/]{har06}.  The spectral energy
distributions (SEDs) of the powerful quasar jets are predominantly characterized as
``optically faint'', with the spectra rising between the optical and X-ray bands.
Current models for this `excess' X-ray emission posit either inverse Compton (IC)
scattering off CMB photons in a (still) relativistic kpc-scale jet or an additional
high-energy synchrotron emitting component. 

In the simplest scenario, such models have diverging predictions at high redshift.
Specifically, we expect a strong redshift dependence in the monochromatic flux
ratio, $f_{X}/f_{r}~\propto~U_{\rm CMB}~\propto~(1+z)^{4}$ for IC/CMB, whereas in
synchrotron models, we expect no such dependence, $f_{X}/f_{r}~\propto~(1+z)^{0}$.
As a first order test of this simple idea, our approach is to study the
highest-redshift relativistic jets. Such jets probe the physics of the earliest
(first $\sim$1 Gyr of the Universe in the quasars studied) actively accreting
supermassive black hole systems and are interesting for other reasons.  For
instance, the ambient medium in these high-redshift galaxies is probably different
\citep[e.g.,][]{dey06} and this may manifest in jets with different morphologies,
increased dissipation, and slower than their lower-redshift counterparts.

\begin{figure}[ht] \plotone{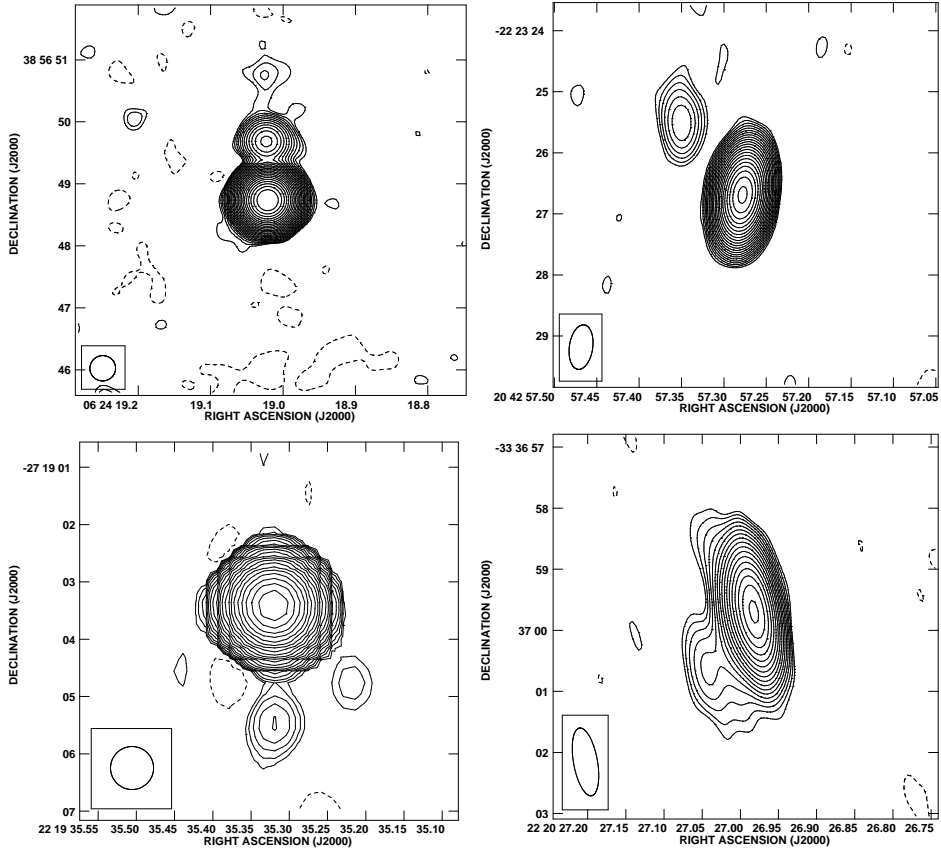}\caption{Examples of newly discovered
arcsecond-scale radio jets from our VLA observations (\S~\ref{sec-vla}). Clockwise
from upper left, the sources are J0624+3856 \citep[$z$=3.469;][]{xu95},
J2042--2223 \citep[$z$=3.630;][]{hoo02}, J2220--3336 \citep[$z$=3.691;][]{hoo02}, and
J2219--2719 \citep[$z$=3.634;][]{hoo02}. The J2219--2719
image is at 1.4 GHz while the rest are at 5 GHz. The beam-sizes are
0.41$''$ $\times$ 0.41$''$, 0.73$''$ $\times$ 0.38$''$ at PA=$-8.2\deg$,
1.13$''$ $\times$ 0.39$''$ at PA=10.2$\deg$, and 0.75$''$ $\times$
0.75$''$ (super-resolved), respectively. The lowest contour levels begin
at 0.125 mJy/bm for all images except for J2219--2719 where it is 0.2
mJy/bm, and increase by factors of $\sqrt{2}$.  \label{fig-1}}\end{figure}

Most $Chandra$ studies of quasar jets have so far targeted known arcsecond-scale
radio jets \citep[e.g.,][]{sam04,mar05}, as most known examples are at $z$
$\stackrel{<}{{}_\sim}$2 \citep{liu02}.  There are currently only two high-$z$
quasars with well-established kpc-scale X-ray jet detections: GB~1508+5714 at
$z$=4.3 \citep{sie03,yua03,che04} and 1745+624 at $z$=3.9 \citep{che06}. They are
observed to have large $f_{X}/f_{r}$ values as expected in the IC/CMB model
\citep{sch02,che04}, although the small number of high-$z$ detections preclude any
definitive statements \citep{kat05,che06}. 

We have therefore carried out a VLA survey in search of new radio jets in a sample of
high-$z$ quasars (\S~\ref{sec-vla}) and new $Chandra$ observations of a small subset
(\S~\ref{sec-cxo}). This contribution presents some results from these observations.
For the redshifts considered, $z$=3.4 to 4.7, 1$''$ corresponds to 7.4 to 6.5 kpc
($H_{0}=70~$km~s$^{-1}$~Mpc$^{-1}$, $\Omega_{\rm M}=0.3$ and $\Omega_{\rm
\Lambda}=0.7$).

\section{Observations of a High-Redshift Quasar Sample}

\subsection{VLA Imaging Survey\label{sec-vla}}

Using NED, we assembled a sample of z$>$3.4 flat-spectrum radio quasars for imaging
with the VLA. We did not aim for our sample to be a complete one as current samples
of lower-z X-ray jets are inhomogenous also. With archival \citep{lee05} and new VLA
observations, we find that radio jets in this redshift range are common with a
$\sim$50$\%$ detection rate \citep[][and in preparation]{che05}. Examples of new
radio jets detected from our observations are shown in Figure~\ref{fig-1}.

\begin{figure}[ht] \plotone{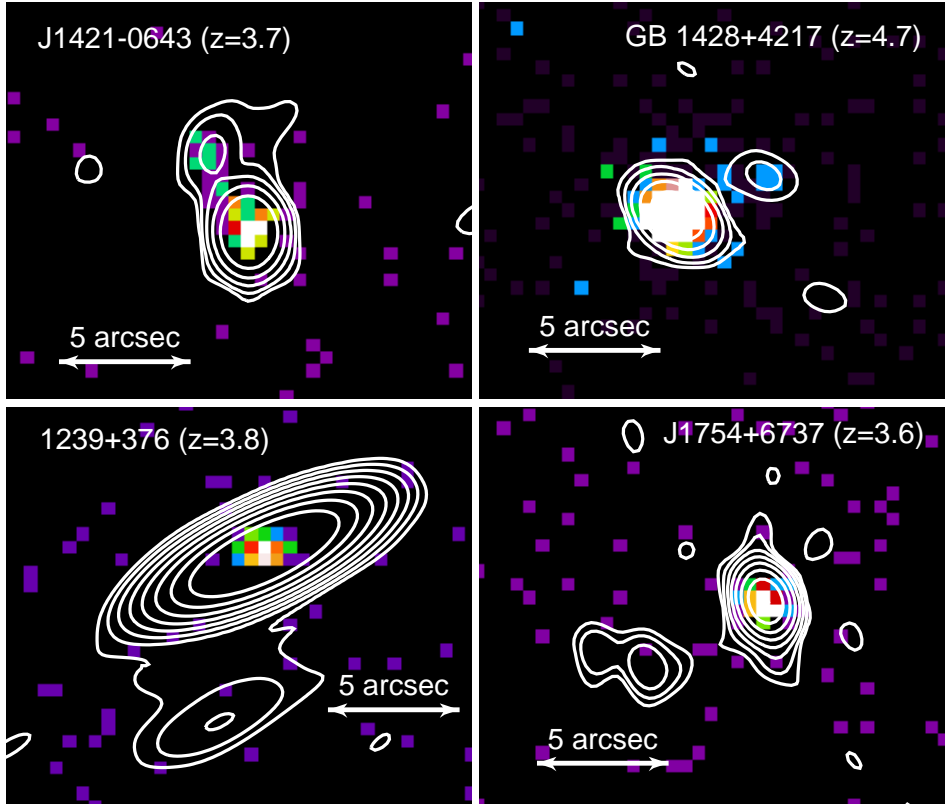}\caption{$Chandra$ X-ray images
(colorscale) with VLA contours overlaid of the four high-$z$ radio jets observed. 
There are X-ray detections of the top two objects but not of the bottom two
(\S~\ref{sec-cxo}). \label{fig-2}}\end{figure}

\subsection{$Chandra$ Observations\label{sec-cxo}}

A small percentage of the radio jets from our radio study (\S~\ref{sec-vla}) are
extended enough ($>$2.5'' long) to study with $Chandra$.  We observed four of them
with short snapshot $Chandra$ observations (Figure~\ref{fig-2}). We detected bright
X-ray counterparts to the jets in the quasars J1421--0643
\citep[$z$=3.689;][]{ell01} and GB~1428+4217 \citep[$z$=4.72;][]{hoo98}; the latter
detection is currently the highest-redshift kpc-scale radio and X-ray jet known. We
did not detect the X-ray counterparts to the radio jets in 1239+376
\citep[z=3.819;][]{ver96} and J1754+6737 \citep[$z$=3.6;][]{vil99}. The 2/4 X-ray
jet detection rate of our high-$z$ sample is comparable to that of lower-$z$ samples
\citep{sam04,mar05}.

\section{Discussion and Summary}

Previous $Chandra$ imaging studies of a number of z$>$4 radio loud quasars do not
reveal significant extended X-ray emission \citep{bas04,lop06}. However, in these
studies, there were no pre-existing information on possible radio structures in the
target objects and any definitive statements regarding the nature of the X-ray
emission mechanism in jets at high-redshifts may be premature. In fact, in one case
where there was evidence of an extended X-ray structure
\citep[J2219--2719;][]{lop06}, our VLA observation revealed a radio counterpart
(Figure~\ref{fig-1}). 

In our approach, we began with a VLA survey of a sample of z$>$3.4 quasars and found
radio jets to be relatively common ($\sim$50$\%$ detection rate). These jets are
quite luminous; with a confident detection of a 1 mJy knot at 1.4 GHz, this
corresponds to luminosities of 1.5 $\times$10$^{42}$ erg s$^{-1}$ ($z$=3.4) to 3.1
$\times$10$^{42}$ erg s$^{-1}$ ($z$=4.7). 

With the radio survey results, we found only a few radio jets to have sufficient
angular extent to be imaged with $Chandra$.  The detection rate of X-ray
counterparts of the high-z radio jets (2/4) is similar to that of lower-z radio jet
samples \citep{sam04,mar05}. The implications of these observations for models of
X-ray emission from large-scale jets will be described in forthcoming publications.

\acknowledgments 

The National Radio Astronomy Observatory is operated by Associated
Universities, Inc. under a cooperative agreement with the National Science
Foundation.  
This research was funded in part by NASA through contract NAS8-39073
(A.~S., D.~E.~H., D.~A.~S.) and $Chandra$ Award Numbers GO7-8114
(C.~C.~C., \L.~S., J.~F.~C.~W., D~.G.) issued by the
$Chandra$ X-Ray Observatory Center, which is operated by the Smithsonian
Astrophysical Observatory for and on behalf of NASA under contract
NAS8-39073. 
Radio astronomy at Brandeis University is supported by the NSF and NASA. 
\L.~S. is supported by MEiN through research project 1-P03D-003-29 from
2005-2008.

\end{document}